# Digital Forensic Investigation of Cloud Storage Services

*Hyunji Chung[a], Jungheum Park[a], Sangjin Lee[a]\*, Chulhoon Kang[b]*

[a]Center for Information Security Technologies (CIST), Korea University, Anam-Dong, Seongbuk-Gu, Seoul, Republic of Korea,
[b]Supreme Prosecutor's Office, Seocho-dong, Seocho-gu, Seoul, Republic of Korea

**Abstract**

The demand for cloud computing is increasing because of the popularity of digital devices and the wide use of the Internet. Among cloud computing services, most consumers use cloud storage services that provide mass storage. This is because these services give them various additional functions as well as storage. It is easy to access cloud storage services using smartphones. With increasing utilization, it is possible for malicious users to abuse cloud storage services. Therefore, a study on digital forensic investigation of cloud storage services is necessary. This paper proposes new procedure for investigating and analyzing the artifacts of all accessible devices, such as Windows, Mac, iPhone, and Android smartphone.

*Keywords*: Digital forensics, Cloud Storage Services, Artifacts, Windows system, Mac system, iPhone, Android.

## 1. Introduction

The development of an information technology infrastructure including faster networks, wider distribution of free software, and advanced virtualization technology has popularized cloud computing. According to Gartner, an IT market research firm, the profits from cloud computing services in 2010 were approximately US $68.3 billion, a 16.6% increase on the previous year, and are expected to reach as much as $148.8 billion in 2014[3] [4].

Cloud computing services can be divided into three kinds, depending on the types of resources provided: SaaS (software as a service), PaaS (platform as a service), and IaaS (infrastructure as a service) [2]. Cloud storage services, a type of IaaS, provide users with virtual space. Designed for business or personal purposes, cloud storage allows users to store data such as documents, images, and music files. Moreover, such services offer not only storage space but also a variety of additional services such as document and image editing, a music and video player, and email-sending capacity. In addition, most cloud storage services allow users to access their cloud storage with a PC or a smartphone to use various services. This access via a device such as a smartphone has helped to spread cloud storage services further.

The popularity of cloud computing has raised questions regarding its security. At the beginning of 2011, a hacker used a cloud service to hack Sony's PlayStation Network[8]. The hacker created an account on Amazon's EC2 service and used the virtual space provided to conduct hacking. This incident showed that there was a need to conduct research on cloud-computing forensics. With the growing popularity of cloud computing, crimes using other types of cloud services are highly likely. In particular, a criminal could leak confidential information from a company by abusing a cloud storage service that allows users to store documents and images and access them through a smartphone.

The hardest aspect of investigating a cloud storage service is that it is difficult to find out what a user did from the moment of subscribing to the service until the end of their use of the service[7]. The log of a cloud server will certainly tell one the history of a user's actions. However, in order to protect clients' personal information, hosting companies are not willing to release information about cloud servers. Nevertheless, it is not impossible to investigate criminal cases involving a cloud storage service, because traces of having used the service are left in a user's device. Webb-Hobson has asserted that although user-made files such as documents, photos, emails, and Internet history are the best digital evidence, such files are not stored in a local device in a cloud computing environment, and so the most important point is where traces of the use of a cloud computing service are stored and how one should analyze them from the perspective of digital forensics[5]. According to Barrett et al., conventional digital forensic methods are insufficient for investigating cloud storage services. In order to conduct a proper investigation, not only conventional computer forensics but also mobile forensics need to be implemented [6]. This paper describes where traces of the use of a cloud storage service exist in local devices (PCs and smartphones) and how these traces should be analyzed.

McClain noted that various traces of using Dropbox can be found in the Windows system. These traces can be located in the installation directory, registry changes on installation, network activity, database files, log files, and uninstallation data[1]. However, although McClain mentioned evidence of these traces, he did not elaborate on what kind of data exist and how they can be utilized to investigate a cloud storage service. In addition, he mentioned that Dropbox can be accessed via a smartphone but did not explain specifically what traces of Dropbox are left there.

Using a cloud storage service leaves traces in local devices. This paper looks into such traces left in PCs and smartphones that can access cloud storage. As the traces left in PCs and smartphones are complementary to each other, all devices that can access a single user's cloud storage must be examined when conducting digital forensics on that storage. In addition, this paper presents methods for collecting and analyzing evidence about a variety of the cloud storage services currently available.

Section 2 of this paper discusses methods for forensic investigation of cloud storage services, and important factors that should be considered in a forensic investigation. Section 3 deals with the traces that are created when a cloud storage service is used with a Windows and Mac system. Section 4 deals with the traces that are left when a cloud storage service is used with two representative smartphone operating systems, namely iOS and Android. Section 5 presents a crime scenario involving a cloud storage service and describes an investigation method. Section 6 presents our conclusions.

## 2. Cloud Storage Services and Digital Forensics

Cloud storage services, a type of IaaS, provide users with storage space. Their use is increasing as they also offer a variety of additional services such as document and image editing, the ability to play music and videos, and email-sending capacity. Most hosting companies provide a certain amount of storage space for free, and a user who wants more space can lease additional storage capacity.

Cloud storage services can be accessed through a Web browser and provide client applications for convenient use. The providers offer client applications for several different platforms to allow people to use their services with diverse devices such as smartphones and tablet PCs.

Cloud storage services provide a range of differentiated services. The artifacts that they create in PCs and smartphones differ between different services, depending on the specific features of the service. In this paper, we have chosen four services, namely Amazon S3, Google Docs, Dropbox, and Evernote. The selection criteria are listed in Table 1.

**Table 1—Selection criteria for the cloud storage services considered in this paper.**

| Common Way to Use | | Type of Service | Public Services |
|---|---|---|---|
| PC (Windows, Mac) | Smartphone (iOS, Android) | | |
| Web browser or Client program | Application | Data storage | **Amazon S3**, **Dropbox**, Sugarsync, and so on |
| | | Office suite + data storage | **Google Docs**, SkyDrive |
| | | Note storage + data storage | **Evernote**, Awesome Note |

The services described in this paper are of three types. The first type provides data storage for the user. To represent this type, we chose Amazon S3, because it is the best-known cloud storage service, and Dropbox, because it is more popular than the alternative Sugarsync [20]. The second type provides both an office suite and data storage. We chose Google Docs, which is better known than SkyDrive [21], to represent this type. The third type gives the user both note storage and data storage. We chose Evernote, which is one of the most popular note-taking applications [22–24], to represent this type.

We recognize that users have many different preferences. In this paper, we have aimed to choose well-known cloud storage services in general.

### 2.1 Procedure for Digital Investigation of Cloud Storage Services

The investigator collects and analyzes data from all devices that a user has used to access a cloud storage service. Such devices include PCs, smartphones, tablet PCs, and PDAs, but this paper covers only PCs and smartphones, which are the mostly widely used devices. A procedure to investigate such devices is shown in Figure 1.

In the case of a Windows and Mac system, the investigator first determines whether it is possible to collect volatile data. If so, the investigator collects the contents of physical memory. Next, if nonvolatile data can be obtained, the investigator gathers data from the Internet history, log files, files, and directories.

In the case of the iOS operating system (iPhone), the investigator can examine backup files stored on a PC, or collect and analyze data used for iTunes. In other words, the investigator can analyze past data backed up on a PC through iTunes or collect data directly from the iPhone for analysis[9]. In the case of an Android phone, data can be acquired after rooting. Rooting is a process that allows users of Android smartphones to attain privileged control. It is an essential process for obtaining data from Android smartphones because one can access the system folder and acquire data only after rooting[10].

The investigator analyzes the data collected from PCs and smartphones as described above, and then checks whether traces of a cloud storage service exist in the collected data. If so, the investigator checks whether information about a user's credentials exists.

If a user's ID and password and any other information that allows access to a user's cloud space are found, a search and seizure warrant should be issued if possible. Otherwise, even if a user's ID and password are found, it will not be possible to log on to the storage service and gather evidence. This is because a user's storage is a private place. Since logging on without a warrant would not be due process, data that was collected without a warrant would not be valid. If the investigator has had a search and seizure warrant issued, the investigator can access to user's storage and collect data in cloud storage. Then, it is possible to analyze data taken from cloud storage and the artifacts that remain in PCs and smartphones. If the investigator has not had a search and seizure warrant issued, it is possible to analyze only the artifacts that remain in PCs and smartphones.

If only the user's ID is found, the investigator checks whether a server belonging to the cloud storage service that the customer has used exists under the same jurisdiction as that of the investigator. If this is the case, a search and seizure warrant should again be issued. After that, the investigator collects the files in the user's cloud storage. If the server is outside the investigator's jurisdiction, the investigator must request international judicial assistance. Cooperation with another nation takes a long time, and in the interval, a malicious customer might delete files in their cloud storage.

In all cases, after the legal proceedings have finished, the investigator analyzes the data that is available from the cloud storage and from artifacts in local devices. The collecting and analyzing data in remote storage is the reason for analyzing the contents of file. To complete the procedure, the investigator writes a report.

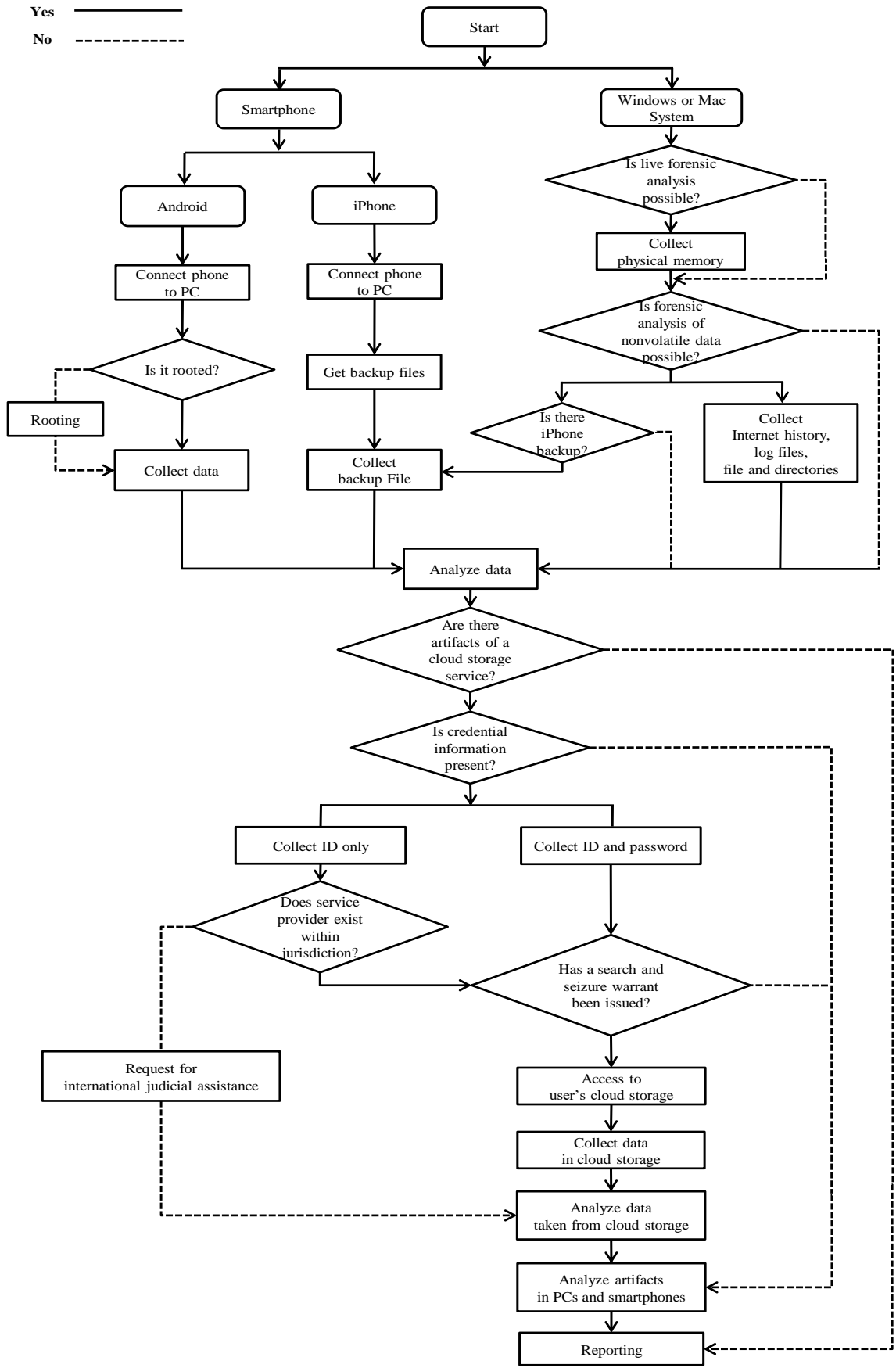

**Figure 1—Procedure for investigation of a cloud storage service.**

## 2.2 Important Factors in an Investigation

This section outlines and provides a rationale for the choice of elements that are prioritized for investigation, among the data collected from PCs and smartphones.

### 2.2.1 Log Files of Web Browsers

As a cloud storage service is basically a Web-based service, it is essential to collect and analyze data regarding Internet history. Here, we look at the files that are left behind when Internet Explorer and Firefox, the most widely used Web browser across the globe, is used. Similar traces can also found in other Web browsers such as Chrome, and Safari. Web browser log files are stored in Profile directory. The Web browser log files consist of cache, history, cookie, and download files. The files exist in the paths listed in Table 2 and Table 3.

**Table 2—Important files and paths (Internet Explorer).**

| OS version | Data | Path |
|---|---|---|
| Windows 2000, XP | Cache | %Profile%\Local Settings\Temporary Internet Files\Content.IE5\index.dat<br>%Profile%\Local Settings\Temporary Internet Files\Content.IE5\<Random>\<All of the files> |
| | History | %Profile%\Local Settings\History\History.IE5\index.dat<br>%Profile%\Local Settings\History\History.IE5\index.dat |
| | Cookie | %Profile%\Cookies\index.dat<br>%Profile%\Cookies\<All of the text file> |
| | Download | Not available |
| Windows Vista, 7 | Cache | %Profile%\AppData\Local\Microsoft\Windows\Temporary Internet Files\Content.IE5\index.dat<br>%Profile%\AppData\Local\Microsoft\Windows\Temporary Internet Files\Content.IE5\<Random>\<All of the files> |
| | History | %Profile%\AppData\Local\Microsoft\Windows\History\History.IE5\index.dat<br>%Profile%\AppData\Local\Microsoft\Windows\History\History.IE5\index.dat |
| | Cookie | %Profile%\AppData\Roaming\Microsoft\Windows\Cookies\index.dat<br>%Profile%\AppData\Roaming\Microsoft\Windows\Cookies\<All of the files> |
| | Download | %Profile%\AppData\Roaming\Microsoft\Windows\IEDownloadHistory\index.dat (From IE 9, only Windows 7) |

**Table 3—Important files and paths (Firefox).**

| OS version | Data | Path |
|---|---|---|
| Mac OS x lion | Cache | /Users/<user name>/Library/Caches/Firefox/Profiles/xxxxxxxx.default/Cache/_CACHE_MAP |
| | History | /Users/<user name>/Library/Application Support/Firefox/Profiles/xxxxxxxx.default/places.sqlite |
| | Cookie | /Users/<user name>/Library/Application Support/Firefox/Profiles/xxxxxxxx.default/cookies.sqlite |
| | Session | /Users/<user name>/Library/Application Support/Firefox/Profiles/xxxxxxxx.default/sessionstore.js |

The cache files include downloaded image files, text files, icons, HTML files, XML files, download URLs, download times, and data sizes. The history files contain URLs that a user has visited, titles of Web pages, the times of visits, and the number of visits. The cookie files store information about hosts, paths, cookie modification times, cookie expiration times, names, and values. The download list includes local paths of downloaded files, download URLs, file sizes, download times, and whether downloads were successful[11]. Through such Web browser files, the investigator can find out about a user's activities such as access or login to a cloud storage service.

On the Windows system, Internet Explorer is the most famous web browser. On the Mac system, Firefox has the highest market share[16]. This paper focuses on log files of Internet Explorer and Firefox.

### 2.2.2 Artifacts of Client Applications in PC

Most cloud storage services provide client applications for convenience. When a client application is installed on a Windows system, traces of it are left in the registry, and log files and database files. Mac system has similar trace except registry files. These are very important files because they contain traces of the use of a cloud storage service. A local PC has log files that contain information on whether or not logins to a cloud storage service were successful, when services were executed and terminated, and when files were synchronized. The examination of log files on a PC provides a broad framework within which one can create a timeline of a user's activities in cloud storage. When a cloud storage service is used, database files are created to manage files and folders in the folders designated for synchronization. These database files contain information about synchronized folders and files on a PC. This information depends on the service provider, but most database files contain records of names of folders and files, their creation times, last modified times, and whether files were deleted.

### 2.2.3 Artifacts in Smartphones

Among the traces left in a smartphone, priority should be given to examining database files, XML files, and plist files. Database files are created to manage files and folders in the folders designated for synchronization. Information about synchronized folders and files on a PC can be obtained from these database files. In addition, information about folders and files accessed through the smartphone can also be found. XML files and plist files must be analyzed because they contain information about the user's account.

### 2.2.4 Physical Memory

Physical memory can contain important information about users, such as IDs and passwords used for logins in a Web browser. It should be collected following the procedure shown in Figure 1 if possible. However, only if system is alive, collecting physical memory is possible. Therefore, this paper doesn't focus on physical memory.

## 3. Artifacts of Cloud Storage Services (Windows and Mac)

Here we describe the artifacts left in the Windows and Mac system after a customer has used a cloud storage service. Table 4 lists some cloud storage services, and the version of the application that we have studied.

**Table 4—Details of services studied in this paper (Windows, Mac, and Smartphone applications).**

| Service name | Details of services | | | |
|---|---|---|---|---|
| | **Windows** | **Mac** | **iOS** | **Android** |
| Amazon S3 | Tested on January, 17, 2012 | Tested on January, 17, 2012 | Cloud Services Manager 2.3 | S3AnywherePro 1.9.1 |
| Dropbox | 1.1.35 | 1.1.43 | 1.4.1 | 1.1.3 |
| Evernote | 4.3.1 | 3.0.1 | 4.0.4 | 3.0.2 |
| Google Docs | Tested on January, 17, 2012 | Tested on January, 17, 2012 | iGoogDocs 1.2 | Google Docs 1.0.10 |

## 3.1 Amazon S3

Amazon S3 is a Web-based cloud storage service and provides various APIs. Many cloud storage services are built using APIs. For example, Dropbox uses Amazon S3's API for data storage [13]. Users can upload, download, open, and delete files[12]. They can access their storage using Windows systems, Mac systems, iPhones, and Android smartphones [18]. Even though Amazon S3 uses SSL by default, temporary Internet files are created.

### 3.1.1 Windows

In Internet Explorer version 8.0, uploading and downloading files and logging users' behavior are possible. Bucket logging is turned off by default. When the user turns it on, log files are delivered to a "bucket log" [17].

When the user downloads and opens a Microsoft Office file on Amazon S3, a file *File name* on s3.amazonaws.com.lnk is created in the path shown in Appendix A. When the user browses the bucket log file, *Log file name*[n].txt is created in another path, also shown in Appendix A. The contents of Table 5 may be helpful for identifying artifacts left by Amazon S3.

**Table 5— Artifacts of Firefox on Windows.**

| Behavior | Type of document | Artifacts |
|---|---|---|
| Download and Open (Not save) | MS Office files (ppt, pptx, doc, docx, xls, xlsx) | *File name* on s3.amazonaws.com.lnk |
| Browsing logging file | txt | *Log file name*[n].txt |

If the user opens the bucket log file in Internet Explorer, a temporary text file is created, such as that shown in Figure 2. The first and second fields in this file are the user's canonical ID and bucket name. The third field is the time at which the user performed the action. The seventh field describes the user's action. The eighth field is the name of the file on which the user acted. The last field is the HTTP user-agent value [17].

```
[Canonical user ID] [Bucket Name] [06/Jan/2012:08:42:20 +0000] [10.186.158.41] [Canonical user ID] [F9ECC0485160EBC5]
[REST.DELETE.OBJECT] [File Name] ["DELETE /Bucket Name/File Name HTTP/1.1"] [204] [-] [-] [42277] [21] [-]["-" ]"[S3Console/0.4]"
```

**Figure 2—Bucket Log File.**

### 3.1.2 Mac

In Firefox version 9.0.1, uploading and downloading files and logging users' behavior are possible. To investigate the use of Amazon S3 with Firefox, general Web browser forensic tools are needed [19]. It is necessary to analyze the URLs that the user has visited and the times at which they were visited. The cache might contain artifacts of files downloaded from Amazon S3. There is no special evidence of Amazon S3 in Firefox.

When the Web browser is closed, the artifacts of Amazon S3 are deleted. However, they can be restored using a tool such as EnCase.

### 3.2 Dropbox

Dropbox is the most frequently used cloud storage service in the world. Whenever a user adds a file to the sync folder, edits a file, or deletes a file, Dropbox automatically syncs the file to its website. The user can access their storage using Windows systems, Mac systems, iPhones, and Android smartphones[13].

#### 3.2.1 Windows

When Dropbox is used on a Windows system, five database files are created in the paths shown in Appendix A. Among them, config.db and filecache.db contain important information. It is easy to identify the contents of these files, because they use the SQLite database format.

Firstly, config.db (Table 6) contains a key named recently_changed3. The value of the key is the names of the five files that the user has edited, copied, moved, and deleted most recently. The file that has been accessed most recently is at the top of the list. From this value, we know which files the user has accessed.

The file config.db also contains the email address for login and the full path in which Dropbox is installed. Even if the investigator does not know the suspect's ID and password, the investigator can access the suspect's cloud storage using config.db. If the investigator collects config.db from the suspect's PC, they can find the suspect's dropbox_path. The investigator installs Dropbox on their PC, copies config.db to that PC, and then places config.db in the same path as the suspect's dropbox_path. By running Dropbox, the investigator can now access the suspect's cloud storage. This method is helpful in an investigation.

**Table 6—config.db.**

| Key | Value |
|---|---|
| dropbox_path | %UserProfile%\AppData\Roaming\Dropbox |
| email | foryou7187@yahoo.co.kr |
| recently_changed3 | (lp1<br>(V41248546 /paper101.doc<br>100<br>tp2<br>a(V41248546 /Digital Forensic of Cloud<br>Ntp3<br>a(V41248546 /Lecture1.pdf<br>Ntp4<br>a(V41248546 /Hello.ppt<br>Ntp5<br>a(V41248546 /snort.pdf<br>Ntp6<br>a. |

Secondly, the times at which a file was created and modified, and the name and path of the synchronized file on the server are stored in filecache.db (Table 7). The format of the time is the Unix time.

**Table 7—filecache.db.**

| server_path | local_filename | local_mtime | local_ctime |
|---|---|---|---|
| 37288970:/Hello | Hello | 1307405626 | 1302685077 |

#### 3.2.2 Mac

When Dropbox is on a Mac system, files are created in the paths shown in Appendix B. Artifacts of Mac are similar to Windows's artifacts except path.

### 3.3 Evernote

Evernote is a well-known storage service that allows a user to save an idea anywhere and anytime. Unlike Dropbox, Evernote synchronizes notes every time they are saved. Users can access their storage using Windows systems, Mac systems, iPhones, and Android smartphones[14].

#### 3.3.1 Windows

When Evernote is used on a Windows system, four folders are created in the paths shown in Appendix A. Among them, the database and logs folders are significant. It is easy to identify the contents of files, because they use the SQLite database format and text.

In the database folder, the files [*userID*].exb and [*userID*].exb.thumbnails exist. [*userID*].exb (Table 8) includes information such as the title of the note, the times at which the note was created and modified, the location where the user created the note, and the type of smartphone operating system that created the note. In addition, information about the attached file can be identified, such as the file name, type, and creation time.

**Table 8—[userID].exb.**

| title | date_created | date_updated | is_deleted | source | latitude | longitude |
|-------|-------------|-------------|-----------|--------|----------|-----------|
| Evernote test | 733700.339618056 | 733700.349456019 | 1 | Mobile.android | 37.321429 | -122.015791 |

The file [*userID*].exb.thumbnails is a combination of PNG files that take a snapshot of the note at every synchronization. The PNG files are listed as shown in Figure 3 (excluding a 24 byte header). By extracting each PNG file, we can know the history of note revisions.

**Figure 3—[*userID*].exb.thumbnails.**

In the logs folder, there are two log files, AppLog_[*Date*].txt (Figure 4) and enclipper_[*Date*].txt. After Evernote has been started, AppLog_[*Date*].txt is created once a day. This file contains authentication information, the account ID, and the times at which the application started and ended. The file enclipper_[*Date*].txt is created once a day, like AppLog_[*Date*].txt. This file includes the time at which the application was started.

**Figure 4— AppLog_[*Date*].txt.**

*3.3.2 Mac*

When Evernote is used on a Mac system, important four files are created in the paths shown in Appendix B. Four files are Evernote.sql, fullscreenThumbnail.png, thumbnail.png, and Evernote.log.

Evernote.sql is SQLite file. Same information can also found in [*userID*].exb (Windows). FullscreenThumbnail.png is an image that takes a snapshot of the note. Thumbnail.png contains text of the note. Evernote.log is same as AppLog_[*Date*].txt (Windows).

## 3.4 Google Docs

Google Docs is a Web-based service that provides flexibility to be productive from one′s desk, on the road, at home and on a mobile phone. When it was launched, Google Docs was classified as a SaaS because it provided users with an application on the Web. Recently, Google Docs has begun to support mobile access by iPhone and Android. Document owners can share and revoke file access at any time. It is possible to upload, download, and edit files. In this respect, it can be considered to belong to the class of cloud storage services[15].

Additionally, even though Google Docs uses SSL by default, Internet temporary files are created.

*3.4.1 Windows*

On the Internet Explorer version 8.0, new Microsoft Office documents, presentations, and spreadsheets can be created. Also, browsing and editing uploaded files is possible. Different types of temporary files are formed in the path (Appendix A), depending on the user's behavior (Table 9). The contents of Table 9 may be helpful for identifying artifacts left by Google Docs.

**Table 9—Artifacts of Internet Explorer on Windows.**

| Behavior | Type of document | Artifacts |
|---|---|---|
| Accessing Google Docs | - | docs_google_com[n].htm |
| Browsing created document | Document | edit[n].htm |
| | Presentation | |
| | Spreadsheet | ccc[n].htm |
| Browsing uploaded document | PDF | viewer[n].htm viewer[n].txt viewer[n].png |
| Editing document | Document | edit[n].htm |
| | ppt | |
| | txt | |
| | Spreadsheet | ccc[n].htm |

After Google Docs has been accessed, docs_google_com[n].htm is created. This file includes list of files in Google Docs. This file starts from docs_google_com[1].htm every day.

When the user browses a Microsoft document or presentation, edit[n].htm is created in the path shown in Appendix A. edit[n].htm contains the contents of the Microsoft document or presentation. In the case of a document, it includes the contents of only one page. When the user browses a Microsoft spreadsheet, ccc[n].htm is created. The contents of the spreadsheet are stored in ccc[n].htm. When the user looks at a pdf file, viewer[n].htm, viewer[n].txt and viewer[n].png are created. The title of the pdf file is saved in viewer[n].htm. The metadata and contents of the pdf file are saved in viewer[n].txt. Each page of the ppt or pdf file is stored in an image in viewer[n].png. When a Microsoft document or a ppt or txt file is edited, edit[n].htm is created. The first page of the document or ppt or txt file is stored in edit[n].htm. When a spreadsheet is edited, ccc[n].htm is created. The contents of the spreadsheet are saved in ccc[n].htm.

When the Web browser is closed, the artifacts of Google Docs are deleted. However, they can be restored using a tool such as EnCase.

*3.4.2 Mac*

On the Firefox version 9.0.1, new Microsoft Office documents, presentations, and spreadsheets can be created. Also, browsing and editing uploaded files is possible. Temporary files are formed in the path (Appendix B), depending on the user's behavior (Table 10). Table 10 may be helpful for identifying artifacts of Google Docs.

**Table 10— Artifacts of Firefox on Mac.**

| Behavior | Type of document | Type of Artifacts |
|---|---|---|
| Browsing uploaded document | ppt | PNG file |
| | pptx | |
| | pdf | |
| Editing document | ppt | HTML file |
| | pptx | |

When the user browses a ppt, pptx, and pdf, PNG files are created in the path shown in Appendix B. Each page of the ppt , pptx, and pdf file is stored in an image in PNG file. The contents of the first page are stored in HTML file. When a ppt and pptx is edited, HTML file is created. The contents of the first page are stored in HTML file. HTML files that include keywords such as "docs" and "id="goog_" are created by Google Docs. The HTML file is verified by a signature <div dir="ltr"> between <body> and </body>. The contents occur after <div dir="ltr">. The contents usually exist between <span> and </span> or between <font> and </font>.

When the Web browser is closed, the artifacts of Google Docs are deleted. However, they can be restored using a tool such as EnCase.

**4. Artifacts of Cloud Storage Services (Smartphones)**

In this section, we demonstrate cloud storage artifacts in two widely used smartphones, namely the iPhone and the Android. The subjects of the experiment were version 4.3.5 of iPhone 4 and version 2.2.2 of Motorola Droid.

**4.1 iOS**

### 4.1.1 Amazon S3

In the paths shown in Appendix C, a plist file and an SQLite database file were generated by running Amazon S3 on an iPhone. Figure 5 shows part of iAwsManager.plist. This file contains the user's name, access key ID, and secret access key. These are used for accessing Amazon S3 on an iPhone.

```
<key>ACCOUNTS</key>
<array>
    <string>HyunjiChung<$$$>Access Key ID<$$$>Secret Acess Key<$$$>NO
    </string>
</array>
```

**Figure 5—iAwsManager.plist.**

iAwsManager.3.0.db includes an important table, named DOWNLOADS (Table 11). This table contains the path, eTag, name and size of downloaded files. It also contains the time at which the user downloaded the file. The field S3BUCKET contains the bucket name.

**Table 11—DOWNLOADS table (iAwsManager.3.0.db).**

| FILENAME | S3KEY | S3BUCKET | FILE_SIZE | DOWNLOAD_DATE |
|---|---|---|---|---|
| Library/downloads/[Forensic.pdf file's eTag].pdf | Forensic.pdf | Hyunjistorage | 8704 | 01/05/12 02:21 PM |

### 4.1.2 Dropbox

In the paths shown in Appendix C, a plist file and two SQLite database files were generated by running Dropbox on an iPhone. Figure 6 shows part of com.getdropbox.Dropbox.plist. This file contains an email address for login, and the first login time using that iPhone.

```
<key>AnalyticsLastUploaded</key>
<date>2011-06-03T06:44:17Z</date>
<key>Dropbox Username</key>
<string>foryou7187@yahoo.co.kr</string>
```

**Figure 6—com.getdropbox.Dropbox.plist.**

The name and path of a folder or file that has been accessed are stored in Dropbox.sqlite, as shown in Table 12. This file also contains the time at which the user browsed the folder or file. As shown in Table 13, Upload.sqlite includes the path and name of the uploaded file, and the time at which it was uploaded. The format of the time is the absolute time.

**Table 12—Dropbox.sqlite.**

| ZPATH | ZLASTVIEWDDATE |
|---|---|
| /folder/Hello.pdf | 329200951.190889 |

**Table 13—Upload.sqlite.**

| ZPATH | ZDATEUPLOADED |
|---|---|
| /folder/Photo 11.6.5 PM 9 02 50.png | 329198703.081349 |

### 4.1.3 Evernote

In the paths shown in Appendix C, various type of files were generated by running Evernote on an iPhone. Firstly, applog.txt is a text file that reports the history of the use of Evernote, as shown in Figure 7. Figure 7 shows the beginning and end of service access, the synchronization time, and connection status (Wi-Fi or 3G). Figure 8 shows the account ID for Evernote in the file com.evernote.iPhone.Evernote.plist.

```
2011-06-03 16:07:04.658 [lvl=2] -[ENAppController _setupSharedLogger]
2011-06-03 16:07:05.343 [lvl=2] -[ENSyncEngine syncIgnoringNetworkPreference:forcingAuthRefresh:]reachability currentReachabilityStatus=WiFi
```

2011-06-03 16:07:05.349 [lvl=2] -[ENSyncEngine _syncStarted] Sync started.

2011-06-03 16:07:05.432 [lvl=2] -[ENSyncEngine _backgroundSync] Contacting Evernote server...

2011-06-03 16:07:08.467 [lvl=2] -[ENSyncEngine(Notebooks) syncNotebooks:] Syncing 2 notebooks...

2011-06-03 16:07:08.661 [lvl=2] -[ENSyncEngine(Notes) updateServerNoteFromLocalNote:] Syncing note 'hallo' [9b79d7a2-3134-453e-a575-ab88a03f8efa]

2011-06-03 16:07:09.462 [lvl=2] -[ENSyncEngine _syncFinished] Sync complete.

**Figure 7—applog.txt.**

<key>username</key>
<string>**dodochung**</string>

**Figure 8—com.evernote.iPhone.Evernote.plist.**

Evernote2.sqlite includes two important tables, ZENLOCALFILE (Table 14) and ZENSERVICEENTITY (Table 15). The times at which the user created, modified, browsed, and deleted a file are stored in the ZLASTACCESSED (ZENLOCALFILE table). The format of the time is the absolute time. And the times at which the user modified a note or file are stored in the ZLASTMODIFIED (ZENLOCALFILE table). ZENSERVICEENTITY contains the title and contents of the note. This table also contains the time at which the note was updated, the type of smartphone operating system that created the note, and the location (latitude and longitude). The format of the time is the absolute time.

All notes and attached files have an index number, assigned sequentially. If there is a number that is not in the index, the note that has that index number has been deleted. When a note is deleted, that note is moved to the recycle bin. A note in the recycle bin includes the time at which it was moved there.

**Table 14—ZENLOCALFILE table (Evernote2.sqlite).**

| ZLASTACCESSED | ZLASTMODIFIED |
|---|---|
| 329205822.2601 | Tue, 07 Jun 2011 06:03:21 GMT |

**Table 15—ZENSERVICEENTITY table (Evernote2.sqlite).**

| ZTITLE | ZDATE UPDATED | ZDATE CREATED | ZDATE DELETED | ZALTITUDE | ZLONGITUDE | ZSOURCE | ZFILENAME SEARCHCONTENT |
|---|---|---|---|---|---|---|---|
| Note1 | 329119401 | 329119401 | -329205827.777706 | 37.321429 | -122.015791 | mobile.android | Photo test.jpg |
| Note2 | 329206645 | 329206158 | 329217190 | 37.590239 | 127.026470 | mobile.iphone | Hello.pdf |

Evernote2.sqlite.md (Figure 9) is an XML file that contains the latest synchronization time.

<key>lastSyncTime</key>
<date>**2011-06-03T07:11:14Z**</date>

**Figure 9— Evernote2.sqlite.md.**

*4.1.4 iGoogDocs*

In the paths shown in Appendix C, a plist file and an html file were generated by running iGoogDocs on an iPhone. Figure 10 shows com.jade.iGoogDocs.plist. This file includes a value for auto login that can be "true" or "false." If the auto login is "true," the plist file contains user account information, including even the password.

<key>password</key>
<string>**googledocspassword**</string>
<key>rememberme</key>
<true/>
<key>username</key>
<string>**localchung@gmail.com**</string>

**Figure 10—com.jade.iGoogDocs.plist.**

In Google Docs for the iPhone, it is possible to create text files. Text files are stored in a folder named "Local Files." The contents of the files are contained in an HTML file. "ios test" in Figure 11 is the text of one file in

"Local Files." Therefore, we can identify the content that the user has written.

```
<html><body><font id='iGoogDocs-Formatted' face='.HelveticaNeueUI' size='17'>ios test</font></br></body></html>
```

**Figure 11—html File(Google Docs).**

## 4.2 Android

### 4.2.1 Amazon S3

In the paths shown in Appendix D, XML file was generated by running Amazon S3 on an Android smartphone. Figure 12 shows s3anywhere.xml. This file contains six important pieces of information.

It contains the name of the bucket that the user accessed. The bucket name is in square brackets. When the user accesses Amazon S3 on an Android smartphone, the remote directory, access key ID, secret access key, and local directory are needed. These are also stored in s3anywhere.xml. This file also contains the last synchronization time. The format of the time is the Unix time. Data between ">" and "<" is encoded in base64. The downloaded file was stored on an external SD card.

Especially, bucket name, access key ID and secret access key are important. This is because these are used for accessing Amazon S3 on an Android smartphone.

```
<string name="s3.remotedir[Bucket name]">Folder's name on Amazon S3</string>

<string name="s3.keyid[Bucket name]"> Access Key ID</string>

<string name="s3.key[Bucket name]"> Secret Access Key</string>

<string name="s3.sync.last.date[Bucket name]">The last synchronization time</string>

<string name="s3.sync.localdir[Bucket name]">Path of local directory</string>
```

**Figure 12—s3anywhere.xml**

### 4.2.2 Dropbox

In the paths shown in Appendix D, two SQLite database files and a log file were generated by running Dropbox on an Android smartphone. These files also contain the email address for login and the full path in which Dropbox is installed. The file prefs.db (Table 16) contains the user's name and email address for login. The file db.db (Table 17) contains the name, size, and time of modification of the uploaded file. The format of the time is the Unix time. The downloaded file was stored in an external SD card.

**Table 16— prefs.db.**

| modified | _display_name | size |
|---|---|---|
| Thu, 17 Mar 2011 00:42:24 +0000 | Forensics.pdf | 47.9KB |

**Table 17—db.db.**

| pref_name | pref_value |
|---|---|
| DISPLAY_NAME | Hyunji Chung |
| COUNTRY | KR |
| EMAIL | foryou7187@yahoo.co.kr |

The file log.txt is a text file that reports the history of execution of Dropbox, as shown in Figure 13. Figure 13 shows times and behaviors, such as success or failure of login attempts, the beginning and end of the service, and file synchronization. The format of the time is the Unix time.

```
1307430069151 com.dropbox.android.DropboxApplication Not authenticated, so authenticating
1307430069376 com.dropbox.android.activity.DropboxBrowser DropboxBrowser starting up
1307516652794 com.dropbox.android.service.DropboxService Dropbox service has been started
1307516652872 com.dropbox.android.service.ServiceBinderDelegate Service is connected
1307516666904 com.dropbox.android.util.FileWatcher Trying to ignore: /mnt/sdcard/dropbox/Photo 3.jpg
1307523391331 com.dropbox.android.activity.lock.LockReceiver LockReceiver received onPause()
1307432400297 com.dropbox.android.taskqueue.MetadataTask Directory changed, going through line-by-line:
                content://com.dropbox.android.Dropbox/metadata/
1307535230918 com.dropbox.android.activity.lock.LockReceiver Received action: android.intent.action.SCREEN_OFF
```

**Figure 13—log.txt.**

*4.2.3 Evernote*

In one of the paths shown in Appendix D, an SQLite database file was generated by running Evernote on an Android smartphone. Table 18 shows part of Evernote.db. This database file includes the title of the note and a flag (is_active) that represents the availability of the note. If the value of the flag is 1, the note is available; otherwise, the note is not available. This file also contains the time at which the note was created and updated, the type of smartphone operating system that created the note, and the location (latitude and longitude). The format of the time is the Unix time. If a note is deleted, that note is moved to the recycle bin. A note in the recycle bin includes a flag (delete is 1) that represents the deletion of the note.

**Table 18—Evernote.db.**

| title | country | created | updated | deleted | is_active | latitude | longtitude | source |
|-------|---------|---------|---------|---------|-----------|----------|-----------|--------|
| Note Test | South Korea | 1307084962000 | 1307426703000 | 0 | 1 | 37.5902 | 127.026 | mobile.iphone |

*4.2.4 Google Docs*

In the paths shown in Appendix D, an SQLite database file and an XML file were generated by running Google Docs on an Android smartphone. DocList.db (Table 19) includes the email address of the account that accessed the smartphone and the last synchronization time. The title of the file, the type of file, and the times of first upload and last modification are stored in DocList.db.

**Table 19—DocList.db.**

| accountHolderName | lastSyncTime | title | kind | creationTime | lastModifiedTime |
|-------------------|--------------|-------|------|--------------|------------------|
| localchung@gmail.com | 1310618312848 | Hello | presentation | 1310618984190 | 1310618984190 |
| abc123@gmail.com | 1310619132115 | Test | spreadsheet | 1310603654097 | 1310613867434 |

There are two important XML files as well as DocList.db. GoodleDriveSharedPreferences.xml includes the email address that the administrator has used. Webview.xml contains the latest email address that has connected to Google Docs. This file is useful in cases where the user has more than two accounts.

## 5. Case Study of a Cloud Storage Service

This section introduces a hypothetical forensic case related to a cloud storage service and describes an investigation of the case using the forensic attributes described in the previous section.

### 5.1 Case Overview

In 2011, Company A found that valuable documents containing designs for a new product had been leaked to a competitor. The file name is "A_design.pdf". In the initial investigation, the prime suspect was a Mr. K, who managed credential files. When data logs across the company's network were checked, no trace remained of any transfer of any relevant file through the network. Mr. K's network traffic gave no clues. However, Mr. K had used a personal computer for business purposes, and also an Android smartphone, so these were target devices. The PC and smartphone were seized for forensic examination. The operating system of the PC was Windows 7 and that of the smartphone was Android.

### 5.2 Objective

The objective of the investigation was to judge whether or not Mr. K had leaked a secret file, by investigating his PC and smartphone.

### 5.3 Method

The investigation started with Mr. K's PC. When the file system, included the unallocated area, was examined, the leaked file was not detected. The log files of the operating system, such as the LKN file, did not provide clues. In the time window when Mr. K might have leaked the file, there were no artifacts resulting from connecting with an external device and accessing a messenger or Web mail service.

While examining the PC, however, the investigator discovered evidence that Dropbox had been installed. The config.db file included the five files that Mr. K had accessed most recently. The five files had random file names. The prime suspect might have changed the file names. To investigate in detail, the investigator has issued a search and seizure warrant. The investigator tried to access the cloud storage. However, the prime suspect refused to reveal his user account information. But, even though the investigator did not know the user ID and password, the investigator was able to access Mr. K's Dropbox storage (see Section 3.2.1), and could see the files as a residue in Dropbox. The investigator collected all files in Dropbox storage. It was discovered that the leaked document was one of the uploaded files in his Dropbox storage. The file name, "A_design.pdf" was changed. "abc.pdf" had the same contents as "A_design.pdf". But no trace of uploading or downloading the files using a smartphone could be identified on the PC.

Hence, Mr. K's smartphone was examined. The investigator connected Android smartphone to PC. After rooting, the investigator collected data related to Dropbox. An artifact of downloading "abc.pdf" existed in log.txt. It was discovered that "abc.pdf" was stored on an external SD card.

## 5.4 Results

In conclusion, he had leaked a confidential file using Dropbox. By analyzing the prime suspect's PC and smartphone together, more precise investigation was possible.

## 6. Discussion and Conclusions

Nowadays, cloud storage services are becoming more well-known. Among cloud computing services, most people uses cloud storage services for managing documents "anytime and anywhere." In particular, the easy accessibility of cloud storage has contributed to the spread of cloud storage services. It is possible for malicious users to abuse cloud storage services, and therefore procedures for forensic investigation of such services are necessary.

Until now, forensic examiners have considered that artifacts of cloud storage services can be examined only in PCs. However, this method is not adequate for investigating cloud storage services, because it fails to provide forensic examiners with further information that does not remain in the PC. This paper has proposed a process model for forensic investigation of cloud storage services, and also described some important elements of an investigation. This paper has described a previously unknown method for forensic analysis of cloud storage services for the Windows, Mac, iOS, and Android operating systems. This methodology is helpful in the investigation of cloud storage services.

**References**

[1] Frank McClain. (2011). Dropbox Forensics.
    (Available online at: http://www.forensicfocus.com/dropbox-forensics)
[2] "Cloud Service". Wikipedia.
    (Available online at : http://en.wikipedia.org/wiki/Cloud_service#Cloud_storage)

[3] Joshua Beil, Bob Egan, Mark Fidelman, Jeffrey Kaplan, Karl Scott, and Joe Tierney. (2010). 2011 Trends Report: Cloud Computing. Focus Research.

[4] Gartner. (2010). Gartner Says Worldwide Cloud Services Market to Surpass $68 Billion in 2010. (Available online at : http://www.gartner.com/it/page.jsp?id=1389313)

[5] Emma Webb Hobson. (2010). Digital Investigations in the Cloud. QinetiQ Digital Investigations Service, Farnborough, UK

[6] Diane Barrett and Gregory Kipper. (2010). Virtualization and Forensics. A Digital Forensic Investigator's Guide to Virtual Environments. 10. Cloud Computing and the Forensic Challenges. Pages 197-209

[7] M.Taylor, J.Haggerty, D.Gresty, and R.Hegarty. (2010). Digital evidence in cloud computing systems. Digital Investigation.

[8] "Amazon cloud service blamed for Sony hacking". Tech.Blorge. (2011). (Available online at : http://tech.blorge.com/Structure:%20/2011/05/16/amazon-cloud-service-blamed-for-sony-hacking/).

[9] iPhone Backup file. (Available online at : http://support.apple.com/kb/ht1766).

[10] Rooting Android. (Available online at : http://android-dls.com/wiki/index.php?title=Rooting_Android).

[11] Jones Keith j and Rohyt Blani. Web browser forensic. Security focus. (Available online at : http://www.securityfocus.com/infocus/1827).

[12] Amazon S3. (Available online at : http://aws.amazon.com/s3/).

[13] Dropbox. (Available online at : https://www.dropbox.com/help/).

[14] Evernote. (Available online at : http://www.evernote.com/).

[15] Google Docs. (Available online at : http://en.wikipedia.org/wiki/Google_Docs).

[16] The highest market share of Mac OS X web browser. (Availble online at : http://web-browsers.findthebest.com/app-question/directory/a/Mac-OS-X/537/What-Mac-OS-X-Web-Browser-has-the-highest-market-share)

[17] Amazon S3 Bucket Logging for audit and to track changes in your bucket. (Available online at : http://www.bucketexplorer.com/documentation/amazon-s3--amazons3-bucket-logging-evidence-of-online-activity.html)

[18] Amazon integrates Apple iPhone, iPad and Android with cloud. (Available online at : http://www.pcworld.idg.com.au/article/370798/amazon_integrates_apple_iphone_ipad_android_cloud/)

[19] Junghoon Oh, Seungbong Lee, Sangjin Lee. Advanced evidence collection and analysis of web browser activity. DFRWS 2011. (Available online at : http://www.dfrws.org/2011/proceedings/12-344.pdf)

[20] Dropbox vs. Sugarsync. (Available online at : http://www.liventerprise.com/compare/Dropbox_vs_Sugarsync/).

[21] Google Docs vs. SkyDrive. (Available online at : http://www.liventerprise.com/compare/SkyDrive_vs_Google_Docs/).

[22] Evernote, Android Market. (Avalilable at : http://itunes.apple.com/us/app/evernote/id281796108?mt=8).

[23] Evernote, iTunes. (Avalilable at : https://market.android.com/details?id=com.evernote&feature=search_result#?t=W251bGwsMSwxLDEsImNvbS5ldmVybm90ZSJd).

[24] Five Best Note-Taking Tools. (Avalilable at : http://lifehacker.com/399556/five-best-note+taking-tools).

## Appendix A. Artifacts of Cloud Storage Service (Windows)

| Service | File system Path | | File name | Details |
|---------|------|--------|-----------|---------|
| | XP | Vista/7 | | |
| Amazon S3 | %UserProfile% \Application Data \Microsoft\Office\Recent | %UserProfile% \Roaming\Microsoft \Office\Recent | *File name* on s3.amazonaws.com.lnk | - MS Office Files that are downloaded and opened |
| | %UserProfile% \Local Settings \Temporary Internet Files \Content.IE5 | %UserProfile%\AppData \Local\Microsoft\Windows \Temporary Internet Files \Content.IE5 | *Log file name*[n].txt | - API that user requests<br>- Time at which user requests API<br>- Name of bucket that accessed Windows system<br>- User's canonical ID |
| Dropbox | %UserProfile% \Application Data \Dropbox | %UserProfile%\AppData \Roaming\Dropbox | config.db | - E-mail address for login<br>- Files that has been accessed most recently(At most five) |
| | | | filecache.db | - Synced file name and path of cloud server<br>- Creation Time<br>- Modification Time |
| Evernote | %UserProfile% \Local Settings \ApplicationData \Evernote\Evernote \Databases | %UserProfile%\AppData \Local\Evernote \Evernote\Databases | userID.exb | - Location that user created note<br>- Flag that represents deletion of note<br>- Type of smartphone operating system<br>- Creation Time<br>- Modification Time<br>- Information about attached file |
| | | | userID.exb.thumbnails | - Combination of PNG files that take a snapshot of note |
| | %UserProfile% \Local Settings \ApplicationData \Evernote\Evernote\Logs | %UserProfile%\AppData \Local\Evernote \Evernote\Logs | AppLog_Date.txt | - Authentication information<br>- Account ID<br>- History of user's behavior |
| | | | enclipper_Date.txt | - Time at which Evernote started |
| Google Docs | %UserProfile% \Local Settings \Temporary Internet Files \Content.IE5 | %UserProfile%\AppData \Local\Microsoft\Windows \Temporary Internet Files \Content.IE5 | docs_google_com[n].htm | - List of files in Google Docs |
| | | | edit[n].htm | - Contents of MS Document and Presentation when browsing it<br>- The first page of MS Document, .ppt, .txt when editing it |
| | | | ecc[n].htm | - Contents of MS Spreadsheet when browsing it<br>- Contents of MS Spreadsheet when editing it |
| | | | viewer[n].xml | - Text of PDF |
| | | | viewer[n].png | - Each page of PDF in image |
| | | | Viewer[n].txt | - The metadata and contents of PDF |

**Appendix B. Artifacts of Cloud Storage Service (Mac)**

| Service | File system Path | File name | Details |
|---------|------------------|-----------|---------|
| Dropbox | /Users/[user name]/.dropbox | config.db | - E-mail address for login<br>- Files that has been accessed most recently(At most five) |
| | | filecache.db | - Synced file name and path of cloud server<br>- Creation Time<br>- Modification Time |
| Evernote | /Users/[user name]/Library/Application Support/Evernote/data | Evernote.sql | - Location that user created note<br>- Flag that represents deletion of note<br>- Type of smartphone operating system<br>- Creation Time<br>- Modification Time<br>- Information about attached file |
| | /Users/[user name]/Library/Application Support /Evernote/data/Contents | fullscreenThumbnail.png | - Full screenshot of note |
| | /Users/[user name]/Library /Application Support/Evernote/logs | thumbnail.png | - Snapshot of content in note |
| | | Evernote.log | - Authentication information<br>- Account ID<br>- History of user's behavior |
| Google Docs | /Users/[user name]/Library/Caches /Firefox/Profiles/[random 8 digits].default/Cache | PNG file | - Each page of uploaded file(ppt, pptx, doc, docx, pdf) in image<br>- Each page of edited file(doc) in image |
| | /Users/[user name]/Library/Caches /Firefox/Profiles/[random 8 digits].default/Cache | HTML file | - Part of contents of doc when browsing it<br>- Part of contents of ppt, pptx when editing it |

**Appendix C. Artifacts of Cloud Storage Service (iOS)**

| Service | File Full Path | File Type | Details |
|---|---|---|---|
| Amazon S3 | Library/Preferences /com.moninnovations.iAwsManager.plist | plist | - User's name<br>- User's access Key ID<br>- User's secret access Key |
| | Document/iAwsManager/iAwsManager.3.0.db | SQLite | - Path, eTag, name and size of downloaded file<br>- Name of bucket that accessed iPhone<br>- Time at which file was downloaded |
| Dropbox | Library/Preferences /com.getdropbox.Dropbox.plist | plist | - E-mail address for login<br>- The first login time |
| | Documents/Dropbox.sqlite | SQLite | - Time at which user browsed folder or file<br>- Name and path of file that user browsed |
| | Documents/Uploads.sqlite | | - Time at which file was uploaded<br>- Name and path of uploaded file |
| Evernote | Documents/www.evernote.com /User/applog.txt | Text | - Beginning and end of service access<br>- Beginning and end of synchronization time<br>- connection status(Wi-Fi, 3G) |
| | Library/Preferences /com.evernote.iPhone.Evernote.plist | plist | - Account ID |
| | Documents/www.evernote.com /User/Evernote2.sqlite | SQLite | - Time at which user created and modified note<br>- Location that user created note<br>- Flag that represents deletion of note<br>- Type of smartphone operating system<br>- Information about attached file<br>- Title and contents of note |
| | Documents/www.evernote.com /User/Evernote2.sqlite.md | XML | - The latest synchronization time |
| Google Docs | Library/Preferences/com.jade.iGoogDocs.plist | plist | - Value for auto login that can be true or false<br>- User's Google Docs ID<br>- User's Google Docs Password (when auto login is true) |
| | Documents/[Title of Document].txt | html | - Contents of created text file |

**Appendix D. Artifacts of Cloud Storage Service (Android)**

| Service | File/Folder Path | File Type | Details |
|---|---|---|---|
| Amazon S3 | /data/data/s3anywherepro/shared_prefs | XML | - Bucket name that accessed Android smartphone<br>- Folder's name on Amazon S3<br>- User's access Key ID<br>- User's secret access Key<br>- The last synchronization time<br>- Path of local directory |
| Dropbox | /data/data/com.dropbox.android/database/prefs.db | SQLite | - User's name<br>- E-mail address for login |
| | /data/data/com.dropbox.android/database/db.db | SQLite | - Name, size, and time of modification of uploaded file |
| | /data/data/com.dropbox.android/files/log.txt | Text | - Success or failure of login attempts<br>- Beginning and end of the service<br>- File synchronization time |
| | /sdcard/dropbox | - | - Existence of file that user downloaded |
| Evernote | /data/data/com.evernote/databases/Evernote.db | SQLite | - Location that user created note<br>- Flag that represents deletion of note<br>- Flag that represents availability of note<br>- Type of smartphone operating system<br>- Creation Time<br>- Modification Time<br>- Time at which note moved to bin |
| | /sdcard/Evernote/notes/content.enml | - | - Contents of note |
| | /sdcard/Evernote/notethumbs | - | - Image file that take a snapshot of note |
| Google Docs | /data/data/com.google.android.apps.docs/databases/DocList.db | SQLite | - Email address of the account that accessed smartphone and the last synchronization time<br>- Title of file<br>- Type of file<br>- Time of the first upload and the last modification |
| | /data/data/com.google.android.apps.docs /shared_prefs/GoogleDriveSharedPreferences.xml | XML | - Email address that the administrator has used |
| | /data/data/com.google.android.apps.docs /shared_prefs/webview.xml | XML | - The lastest email address that has connected to Google Docs |